\documentclass[12pt]{article}
\usepackage{amsmath}
\usepackage{amssymb}
\usepackage{graphicx}
\usepackage{epsfig}
\usepackage{subfigure}
\usepackage{url}
\usepackage[small]{caption}
\begin{document}
\baselineskip 0.6cm

\def\simgt{\mathrel{\lower2.5pt\vbox{\lineskip=0pt\baselineskip=0pt
           \hbox{$>$}\hbox{$\sim$}}}}
\def\simlt{\mathrel{\lower2.5pt\vbox{\lineskip=0pt\baselineskip=0pt
           \hbox{$<$}\hbox{$\sim$}}}}
\def\simprop{\mathrel{\lower3.0pt\vbox{\lineskip=1.0pt\baselineskip=0pt
             \hbox{$\propto$}\hbox{$\sim$}}}}
\def\bra#1{\left< #1 \right|}
\def\ket#1{\left| #1 \right>}
\def\inner#1#2{\left< #1 | #2 \right>}

\begin{titlepage}

\begin{flushright}
MIT-CTP-4452 \\
UCB-PTH-13/03 \\
\end{flushright}

\vskip 1.0cm

\begin{center}
{\Large \bf Low Energy Description of Quantum Gravity \\ and Complementarity}

\vskip 0.7cm

{\large Yasunori Nomura, Jaime Varela, and Sean J. Weinberg}

\vskip 0.4cm

{\it Center for Theoretical Physics, Laboratory for Nuclear Science, 
     and Department of Physics, \\
     Massachusetts Institute of Technology, Cambridge, MA 02139, USA} \\

\vskip 0.2cm

{\it Berkeley Center for Theoretical Physics, Department of Physics, \\
     and Theoretical Physics Group, Lawrence Berkeley National Laboratory, \\
     University of California, Berkeley, CA 94720, USA} \\

\vskip 0.8cm

\abstract{We consider a framework in which low energy dynamics of 
 quantum gravity is described preserving locality, and yet taking into 
 account the effects that are not captured by the naive global spacetime 
 picture, e.g.\ those associated with black hole complementarity.  Our 
 framework employs a ``special relativistic'' description of gravity; 
 specifically, gravity is treated as a force measured by the observer 
 tied to the coordinate system associated with a freely falling local 
 Lorentz frame.  We identify, in simple cases, regions of spacetime in 
 which low energy local descriptions are applicable as viewed from the 
 freely falling frame; in particular, we identify a surface called the 
 gravitational observer horizon on which the local proper acceleration 
 measured in the observer's coordinates becomes the cutoff (string) 
 scale.  This allows for separating between the ``low-energy'' local 
 physics and ``trans-Planckian'' intrinsically quantum gravitational 
 (stringy) physics, and allows for developing physical pictures of the 
 origins of various effects.  We explore the structure of the Hilbert 
 space in which the proposed scheme is realized in a simple manner, 
 and classify its elements according to certain horizons they possess. 
 We also discuss implications of our framework on the firewall problem. 
 We conjecture that the complementarity picture may persist due to 
 properties of trans-Planckian physics.}

\end{center}
\end{titlepage}

\section{Introduction}
\label{sec:intro}

In the past few decades, it has become increasingly clear that quantum 
theory of gravity will not be built on a simple global spacetime picture of 
classical general relativity.  Quantum mechanics requires a large deviation 
from the simple global spacetime picture even at long distances---distances 
much larger than the fundamental scale $l_*$, which is expected to be 
close to the 4-dimensional Planck length $l_{\rm Pl} \simeq 1.62 \times 
10^{-35}~{\rm m}$.  General relativity must arise as an effective 
theory---not in the simplest Wilsonian sense---describing observations 
performed by classical observers.

Historically, the first hint of this has come from studying black 
holes.  The standard local formulation of quantum gravity leads to 
inconsistency when describing a process in which an object falls into 
a black hole that eventually evaporates, since it may employ a class of 
equal time hypersurfaces (called nice slices) on which quantum information 
is duplicated~\cite{Preskill:1992tc}.  In the early 90's, a remarkable 
suggestion to avoid this difficulty---called the complementarity 
picture---has been made~\cite{Susskind:1993if,Stephens:1993an}:\ 
the apparent cloning of the information occurring in black hole physics 
implies that the internal spacetime and horizon/Hawking radiation 
degrees of freedom appearing in different, i.e.\ infalling and distant, 
descriptions are not independent.  This clearly signals a breakdown of 
the naive global spacetime picture of general relativity, and forces 
us to develop a new low energy theory of quantum gravity in which 
locality is preserved (if there exists such a formulation).

In this letter, building on earlier suggestions in 
Refs.~\cite{Nomura:2011rb,Nomura:2011dt}, we propose an explicit 
framework in which low energy dynamics of quantum gravity is described 
preserving locality, and yet taking into account the effects that are 
not captured by the naive global spacetime picture.  We introduce an 
explicit coordinate system associated with a freely falling reference 
frame, which we call the observer-centric coordinate system, that 
allows for a ``special relativistic'' description of gravity, i.e.\ 
treating gravity {\it as a force} measured by the observer tied to 
this coordinate system.  This allows us to identify, in simple cases, 
boundaries of spacetime where the local description of the system 
breaks down, which we call the observer horizon.  We propose a specific 
Hilbert space, which we refer to as the covariant Hilbert space for quantum 
gravity, in which the proposed scheme is realized in a simple manner. 
We also discuss possible implications of this framework for the firewall 
problem~\cite{Almheiri:2012rt}, i.e.\ how it allows for keeping the 
basic hypotheses of complementarity:\ (1) unitarity of quantum mechanics, 
(2) the validity of semi-classical descriptions of physics outside 
the stretched horizon, and (3) the equivalence principle (no ``drama'' 
at the horizon for an infalling object), under certain assumptions 
on microscopic physics at scales above $M_* \equiv 1/l_*$.%
\footnote{The fundamental scale $M_*$ is related to the Planck scale 
 $M_{\rm Pl} \equiv 1/l_{\rm Pl}$ by $M_{\rm Pl}^2 \sim N M_*^2$ in four 
 dimensions, where $N$ is the number of species existing below $M_*$.}
More detailed discussions on the framework described here, including 
the basic philosophy motivating it, will be presented in the upcoming 
paper~\cite{NVW}.

In this letter we limit our discussions to the case of four spacetime 
dimensions, but the extension to other dimensions is straightforward. 
We take the Schr\"{o}dinger picture throughout, and we work with a metric 
signature that is mostly positive.

\section{Covariant Hilbert Space for Quantum Gravity}
\label{sec:QHS}

Our construction is based on a series of hypotheses.  More detailed 
descriptions, as well as motivations, of these hypotheses will be given 
in Ref.~\cite{NVW} (see also~\cite{Nomura:2011rb}).  Here we simply 
list them without much elaboration.

We postulate
\begin{itemize}
\item[(i)] A Hamiltonian formalism exists that describes a quantum 
 mechanical system with gravity.  Since a system with gravity in general 
 has constraints, we consider the constrained Hamiltonian formalism 
 developed by Dirac~\cite{Dirac-book}.
\item[(ii)] There is a way to restrict Hilbert space (e.g.\ fix intrinsically 
 stringy gauge redundancies) in such a way that dynamics defined on it 
 is local in spacetime at length scales larger than $l_*$.  In other words, 
 there is a way to formulate a theory such that ``intrinsically quantum 
 gravitational'' (stringy) effects decouple at distances larger than 
 $l_*$ (the string scale).
\item[(iii)] The desired local description is obtained by restricting 
 the Hilbert space such that an element represents either an appropriately 
 restricted region of a spacetime hypersurface (when it allows for a 
 spacetime interpretation) or an intrinsically quantum gravitational state 
 (when it does not).  In particular, the former can be taken to represent 
 a state of physical degrees of freedom on {\it a portion of} the past 
 light cone of a fixed reference point $p_0$.
\end{itemize}
A main motivation for the last hypothesis is that it seems to constitute 
the minimal deviation from the standard general relativistic view of 
spacetime, needed to address the issue of information cloning in the 
existence of a horizon.  The use of a light cone is also motivated to 
make the causal structure manifest; the hypersurface represented by 
a state corresponds to the spacetime region from which a hypothetical 
observer at $p_0$ can obtain light ray signals.  (The possibility 
of using a spacelike hypersurface will be mentioned later.)

We argue that the desired description is obtained by suitably dropping 
some of the constraints needed to reduce the Hilbert space to that of 
the physical states:
\begin{equation}
  {\cal P}^\mu({\bf x}) \ket{\Psi} = 0,
\label{eq:GR-const}
\end{equation}
where $\mu = 0,\cdots,3$, and ${\bf x}$ are the coordinates parameterizing 
a hypersurface on which the states are defined.  (For more detailed 
discussions, see Ref.~\cite{NVW}.)  Note that $\ket{\Psi}$ represents 
a quantum state for the {\it entire} system, including possible degrees 
of freedom associated with the boundaries of space, which may be located 
at infinity.  Now, a natural way to define locality is through the 
structure of the Hamiltonian.  However, if we define the Hamiltonian 
operator (which is a linear combination of ${\cal P}^\mu({\bf x})$'s) 
on Hilbert space ${\cal H}_{\rm phys}$ spanned by the independent 
physical states $\ket{\Psi}$, then it is simply zero.  Furthermore, 
it is in general not possible to take a basis in ${\cal H}_{\rm phys}$ 
such that all of its elements represent well-defined semi-classical 
spacetimes as they are generically in superposition states.%
\footnote{Because the quantum state we consider here, $\ket{\Psi}$, is 
 the state representing the entire system including clock degrees of 
 freedom (as opposed to relative states $\ket{\psi_i}$ which may evolve 
 in time), it satisfies all the constraints in Eq.~(1), including the 
 Hamiltonian constraint.  This makes $\ket{\Psi}$ a superposition of 
 terms representing semi-classical spacetimes because it takes the 
 form of $\ket{\Psi} = \sum_i \ket{i} \ket{\psi_i}$, where $\ket{i}$ 
 and $\ket{\psi_i}$ represent the clock degrees of freedom and the rest 
 of the system, respectively.}
This precludes us from labeling the elements of ${\cal H}_{\rm phys}$ 
according to physical configurations in spacetime, since they do 
not even have well-defined spacetimes.  In particular, in the Hilbert 
space ${\cal H}_{\rm phys}$ spanned by physical (gauge invariant) states 
$\ket{\Psi}$, local operators---or the concept of locality itself---cannot 
be defined in general.

These issues can be addressed if we consider a Hilbert space larger 
than ${\cal H}_{\rm phys}$ by appropriately dropping some of the 
constraints (which then must be imposed later as the ``dynamics'' of 
the system).  Specifically, consider a (hypothetical) reference point 
$p_0$ at some ${\bf x}$ and a local Lorentz frame elected there.  We 
may then change the basis of constraint operators ${\cal P}^\mu({\bf x})$ 
(by taking their linear combinations) so that it minimizes the number 
of constraints corresponding to transformations affecting the local 
Lorentz frame.  This leads to $10$ constraint operators, $H$, $P_i$, 
$J_{[ij]}$, and $K_i$ ($i = 1,2,3$), associated with the change of the 
local Lorentz frame.  These operators obey the standard Poincar\'{e} 
algebra.  (The set of operators determined in this way is not unique, 
and each choice corresponds to adopting different, e.g.\ null or 
spacelike, quantization.)

We now postulate
\begin{itemize}
\item[(iv)] By dropping the constraints related to the changes of the local 
 Lorentz frame
 \begin{equation}
    H \ket{\Psi} = P_i \ket{\Psi} = J_{[ij]} \ket{\Psi} = K_i \ket{\Psi} = 0,
 \label{eq:Poincare}
 \end{equation}
 we obtain a Hilbert space ${\cal H}_{\rm QG}$ larger than 
 ${\cal H}_{\rm phys}$.  The elements of ${\cal H}$---the subspace 
 of ${\cal H}_{\rm QG}$ allowing for a spacetime interpretation---can 
 then be labeled by physical configurations in spacetime hypersurfaces 
 (together with possible other labels such as spins of particles); 
 in other words, we can take a basis of ${\cal H}$ such that all the 
 basis states have well-defined semi-classical spacetimes.  Physics 
 defined on this space is local in the bulk of spacetime.
\end{itemize}
In particular, we assume that we can take specific linear combinations 
of the constraint operators ${\cal P}^\mu({\bf x})$ such that the 
appropriate basis states of ${\cal H}$ represent the configurations 
of physical degrees of freedom on (portions of) the past light cone 
of $p_0$.  We then call the corresponding enlarged Hilbert space 
${\cal H}_{\rm QG}$ the {\it covariant Hilbert space for quantum 
gravity}.%
\footnote{The physical Hilbert space, ${\cal H}_{\rm phys}$, is 
 a subspace of ${\cal H}_{\rm QG}$.  As such, any gauge-invariant 
 (constrained) state, i.e.\ an element of ${\cal H}_{\rm phys}$, can 
 be expanded as a superposition of elements in ${\cal H}_{\rm QG}$ 
 in the ``locality basis'' that can be determined by the structure 
 of the Hamiltonian defined in this enlarged Hilbert space.}

The Hamiltonian defined on ${\cal H}_{\rm QG}$ represents local physics 
on the past light cone of $p_0$ within a boundary, which we will explicitly 
determine in simple cases below.  (Here we are considering each component 
state, i.e.\ a basis state in ${\cal H}$ in the basis given in (iv). 
The full quantum state is in general a superposition of these and other 
states.)  This Hamiltonian is not {\it manifestly} local, since the 
constraints associated with the coordinate transformations on the 
past light cone of $p_0$ are still imposed on ${\cal H} \subset 
{\cal H}_{\rm QG}$.  In other words, the elements of ${\cal H}$ 
represent physical states obtained after solving Einstein's equation 
on the light cone.  To recover a manifest locality of the Hamiltonian, 
we need to introduce appropriate metric degrees of freedom on the light 
cone and drop the corresponding constraints from the definition of 
the Hilbert space.  We assert that the resulting Hamiltonian is then 
manifestly local in the bulk of spacetime (but not at the boundary). 
In the rest of the letter, we do not bother with this last step and 
focus our attention on ${\cal H}_{\rm QG}$, which is enough to make 
{\it physics} local in the bulk (in the sense that there exists an 
equivalent, though more redundant, description in which the Hamiltonian 
takes a manifestly local form).%
\footnote{The commutation relations among field operators may contain 
 apparent non-local terms associated with null quantization, which 
 arise from the fact that massless particles can propagate along the 
 light cone. \label{ft:null}}

The Hilbert space ${\cal H}_{\rm QG}$ is the relevant Hilbert space when 
we discuss ``evolution'' of a system with gravity.  It is true that a 
physical state of the {\it entire} system must obey all the constraints, 
including those in Eq.~(\ref{eq:Poincare}), and thus satisfies
\begin{equation}
  \frac{d}{dt} \ket{\Psi} = 0,
\label{eq:no-time-ev}
\end{equation}
i.e.\ $\ket{\Psi}$ is static.  However, in $\ket{\Psi}$ we can identify 
a (small) subsystem as the ``clock'' degrees of freedom, and rewrite the 
entanglement of these degrees of freedom---represented e.g.\ by a set of 
states $\ket{i}$---with the rest of the degrees of freedom---represented 
e.g.\ by a set of states $\ket{\psi_i}$---in the standard form of 
Schr\"{o}dinger time evolution of a state $\ket{\psi_i}$, where $i$ 
plays the role of time~\cite{DeWitt:1967yk}.  (In the Minkowski space, 
we are doing this operation implicitly by identifying boundary degrees 
of freedom at infinity as the clock degrees of freedom; this is why 
we can consider time evolution, or $S$ matrix, in Minkowski space without 
explicitly being bothered by the clock degrees of freedom.)  We may 
then view ${\cal H}_{\rm QG}$ as the Hilbert space in which $\ket{\psi(t)} 
\equiv \ket{\psi_i}$ evolves unitarily according to the ``derived'' 
Hamiltonian, which in general depends on the choice of the clock 
degrees of freedom.%
\footnote{In order for this operation to give well-defined time evolution 
 of $\ket{\psi_i}$ by an ordered Hamiltonian at a macroscopic level, 
 the state $\ket{\Psi}$ must be in a special low coarse-grained 
 entropy state, at least in branches relevant for the clock degrees 
 of freedom.  In a real cosmological situation, when $\ket{\Psi}$ 
 represents the entire ``multiverse state,'' this leads to a set of 
 conditions which the Hamiltonian $H$ defined on ${\cal H}_{\rm QG}$ 
 must satisfy~\cite{Nomura:2012zb}.}
(Note that $\ket{\psi_i}$ are no longer zero eigenvalue eigenstates 
of $H$, $P_i$, $J_{[ij]}$, or $K_i$, in general.)  Furthermore, 
complementarity can be viewed as a relation between different low 
energy descriptions corresponding to different choices of clocks 
separated beyond each other's horizon, which are obtained after 
a suitable action of $H$, $P_i$, $J_{[ij]}$, or $K_i$ to put the 
clock in the bulk of spacetime in each description.  From this 
perspective, $\ket{\Psi}$ serves the role of a generating function 
from which physical predictions can be derived by identifying the 
clock degrees of freedom and extracting their entanglement with 
the rest.  For more detailed discussions on this construction, 
see Ref.~\cite{NVW} (also~\cite{Nomura:2011rb}).

We note that while our framework allows for formally writing down the 
Hamiltonian applicable at length scales larger than $l_*$, this is 
not directly useful in calculating the effect of dynamical spacetime 
or the result of a reference frame change, since they depend on unknown 
dynamics of degrees of freedom at the boundaries of space.  This problem 
may be largely bypassed if we are interested only in a coarse-grained 
description of the system, by employing a certain correspondence principle 
which we may call the complementarity hypothesis~\cite{Nomura:2012cx}---we 
can then use a combination of quantum theory and classical general 
relativity to obtain a coarse-grained description of the evolution 
of the system.  An advantage of our framework in doing this is that 
it clearly separates between the ``low-energy'' local physics and 
``trans-Planckian'' intrinsically quantum gravitational (stringy) 
physics, so it allows for developing clear physical pictures of the 
origins of various effects.  To obtain a complete dynamical theory, 
however, we would need to formulate the theory applicable above 
$M_*$---presumably string theory---along the lines described here. 
This is beyond the scope of the present work.

The structure of the covariant Hilbert space takes the form
\begin{equation}
  {\cal H}_{\rm QG} = {\cal H} \oplus {\cal H}_{\rm sing}.
\label{eq:H_QG}
\end{equation}
Here, ${\cal H}$ is spanned by all the possible physical configurations 
realized on (portions of) the past light cone of $p_0$ {\it as viewed 
from a local Lorentz frame at $p_0$}, while ${\cal H}_{\rm sing}$ 
contains intrinsically quantum mechanical states that do not allow for 
a spacetime interpretation (the states relevant when $p_0$ hits a 
spacetime singularity), where ${\rm dim}\, {\cal H}_{\rm sing} = 
\infty$~\cite{Nomura:2011rb}.  How do we define physical configurations 
``as viewed from a local Lorentz frame at $p_0$''?  Where is the boundary 
of space that determines the relevant portion of the light cone for each 
element of ${\cal H}$?  In the next section, we address these questions 
and provide an explicit prescription to specify elements of ${\cal H}$ 
which is applicable in simple cases.  We also discuss a global structure 
of ${\cal H}$, based on a certain classification scheme for the elements.

\section{Defining Boundaries and Classifying the States}
\label{sec:boundary}

We now focus on ${\cal H}$ and identify a spacetime region (in particular, 
a region on an ``equal-time'' hypersurface) represented by its element. 
We discuss how independent quantum states comprising ${\cal H}$ are 
specified, and classify them into elements of ${\cal H}_{\partial{\cal M}}$'s, 
the subsets of ${\cal H}$ labeled by ``horizons'' possessed by the states.

\subsection{Observer-centric coordinates}
\label{subsec:coordinates}

We first introduce a useful coordinate system to describe our construction. 
Let us choose a fixed spacetime point $p_0$ in a fixed spacetime 
background.  We consider that an element of ${\cal H}$ represents physical 
configurations of dynamical degrees of freedom and their conjugate 
momenta on the past light cone of $p_0$, which we call $L_{p_0}$.  In 
general, the elements of ${\cal H}$ are labeled by a set of quantum 
numbers (i.e.\ the response to a set of quantum operators), and in 
Section~\ref{subsec:classify} we will discuss how many independent 
such quantum states exist in full quantum gravity.  For now, however, 
it is sufficient to keep in mind that the state is specified by the 
response to the operators defined on $L_{p_0}$.

Now, consider a timelike geodesic $p(\tau)$ which passes through $p_0$ 
at $\tau = 0$: $p(0) = p_0$.  We take $\tau$ to be the proper time 
measured at $p$.  A set of local Lorentz frames elected along $p(\tau)$ 
corresponding to a freely falling frame can then be uniquely determined 
by specifying spacetime location $q^\mu$ and proper velocity $v^i$ of 
$p$ at $\tau = 0$
\begin{equation}
  x_{p(0)}^\mu = q^\mu,
\qquad
  \frac{d x_{p(\tau)}^i}{d\tau} \bigg|_{\tau=0} = v^i,
\label{eq:p-spec}
\end{equation}
as well as 3 Euler angles $\alpha^{[ij]}$ that determine the orientation 
of the coordinate axes, where $i = 1,2,3$.  This is because all the 
axes of the local Lorentz frames along $p(\tau)$ can be obtained by 
parallel transporting the axes at $p(0)$.

\begin{figure}[t]
\begin{center}
  \includegraphics[width=10cm]{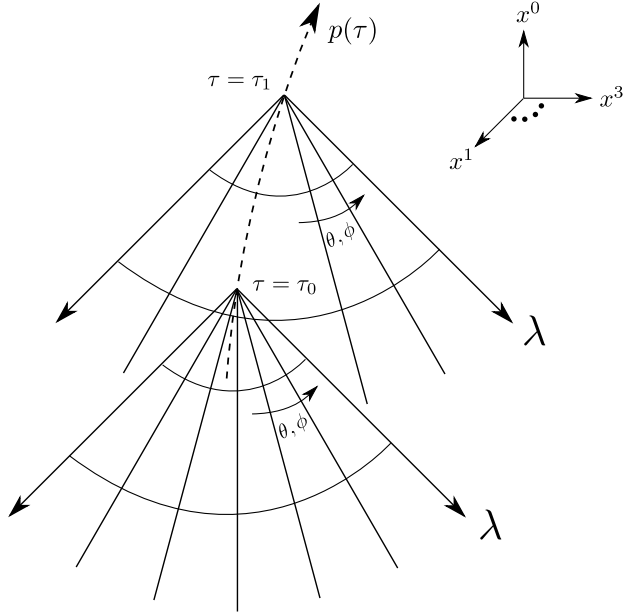}
\caption{A schematic depiction of the observer-centric coordinate system.}
\label{fig:obs-coord}
\end{center}
\end{figure}

We now introduce angular coordinates $(\theta,\phi)$ at each $\tau$ 
which coincide with the angular variables of the spherical coordinate 
system of the local Lorentz frame in an infinitesimally small neighborhood 
of $p(\tau)$.  We then define the ``radial'' coordinate $\lambda$ for 
fixed $\tau, \theta, \phi$ as the affine parameter associated with the 
light ray emitted toward the past from $p(\tau)$ in the direction of 
$(\theta,\phi)$.  The origin and normalization of $\lambda$ are taken 
so that the values of $\lambda$ agree with those of the radial coordinate 
of the local Lorentz frame in an infinitesimally small neighborhood of 
$p(\tau)$.  We perform this procedure in an inextendible spacetime; for 
example, we do not terminate the light ray at a coordinate singularity. 
This process allows us to introduce the coordinate system, which we 
call the {\it observer-centric coordinate system}.  It has 4~coordinates 
$\tau$, $\lambda$, $\theta$, and $\phi$, depicted schematically in 
Fig.~\ref{fig:obs-coord}, and provides a reference frame from which 
physics is described. 
Note that a hypersurface with constant $\tau$ corresponds to the past 
light cone of $p(\tau)$, which is a null, rather than spacelike, 
hypersurface.  To describe a state, we need this coordinate system 
only in an infinitesimally small neighborhood of the $\tau = 0$ 
hypersurface.  The reason why we need the neighborhood is that some 
phase space variables involve the $\tau$ derivative of quantum fields 
at $\tau = 0$.

We describe a quantum state, e.g.\ the configuration of matter on the 
``equal time'' (null) hypersurface, using the observer-centric coordinate 
system throughout the evolution of the system.  The introduction of 
this ``absolute coordinate system'' allows us to view gravity as a force 
measured in these coordinates---the motion of a particle of mass $m$ 
under the influence of gravity can be expressed as $m \ddot{\chi} = F$, 
where $\chi = (\lambda, \theta, \phi)$ and the dot represents a $\tau$ 
derivative.

For a given spacetime, we may convert a coordinate system $x^\mu$ to 
the observer-centric one once a local Lorentz frame is elected.  For 
this purpose, we regard $x^\mu$ to be functions of the observer-centric 
coordinates, $x^\mu(\tau, \lambda, \theta, \phi)$, and derive equations 
that allow us to solve these functions.  Note that the form of these 
functions depends on the choice of the local Lorentz frame, $(q^\mu, 
v^i, \alpha^{[ij]})$.

\subsection{Gravitational observer horizon}
\label{subsec:observer-hor}

In general, an element of ${\cal H}$ represents only a portion of 
$L_{p_0}$.  Specifically, a past-directed light ray emitted from $p_0$ 
will hit a point beyond which the semi-classical description of spacetime 
is not applicable.  The collection of these points forms a two-dimensional 
surface
\begin{equation}
  \lambda = \lambda_{\rm obs}(\theta,\phi),
\label{eq:obs-horizon}
\end{equation}
which we call the {\it gravitational observer horizon}, or the observer 
horizon for short.  In general, we expect that this surface is determined 
by some condition which indicates that the intrinsically quantum 
gravitational physics becomes important there.  In some simple cases, 
however, we may be able to state the condition more explicitly.

Consider a spacetime trajectory of a point with constant $(\lambda, 
\theta, \phi)$ in the infinitesimal vicinity of $L_{p_0}$.  Its proper 
velocity is given by
\begin{equation}
  u^\mu = \frac{\frac{\partial x^\mu}{\partial \tau}}
    {\sqrt{-g_{\mu\nu} \frac{\partial x^\mu}{\partial \tau} 
    \frac{\partial x^\nu}{\partial \tau}}},
\label{eq:proper-vel}
\end{equation}
while the local proper acceleration by
\begin{equation}
  a^\mu = u^\nu \nabla_\nu u^\mu.
\label{eq:proper-acc}
\end{equation}
Here, $x^\mu$ is an arbitrary coordinate system.  $a^\mu(\tau,\lambda=0,
\theta,\phi) = 0$ since $p(\tau)$ is a geodesic, but if $\lambda >0$, 
a trajectory of constant $(\lambda, \theta, \phi)$ need not be a geodesic 
so we may have $a^\mu(\tau, \lambda, \theta, \phi) \neq 0$.  $a^\mu$ has 
dimensions of energy in natural units $\hbar = c = 1$.  Note that $u^\mu$ 
is timelike while $a^\mu$ is spacelike (or zero) within a (coordinate) 
horizon $g_{\tau\tau} = g_{\mu\nu} (\partial x^\mu/\partial \tau) 
(\partial x^\nu/\partial \tau) = 0$, where these vectors diverge.

In general, special behaviors of these quantities, e.g.\ $g_{\tau\tau} 
\rightarrow 0$ and $a^\mu \rightarrow \infty$, may be merely coordinate 
artifacts.  We claim, however, that when the system under consideration 
is static, i.e.\ when the spacetime admits a timelike Killing vector 
$k^\mu$ and when the geodesic, $p(\tau)$, is approximately along this 
vector ($ dp^\mu(\tau)/d\tau \simprop k^\mu$), then the surface on 
which the magnitude of the local proper acceleration vector $a^\mu$ becomes 
the cutoff scale $M_*$ signals the breakdown of the semi-classical 
description, giving the surface $\lambda = \lambda_{\rm obs}(\theta,\phi)$. 
Namely, in a static situation, the semi-classical picture is applicable 
only on a portion of $L_{p_0}$ in which
\begin{equation}
  A \equiv \sqrt{a^\mu a_\mu} \simlt M_*.
\label{eq:eff-theory}
\end{equation}
This is a natural criterion given that $a^\mu$ measures acceleration 
relative to a free-fall.  It can be interpreted as the condition that 
the gravitational acceleration measured from the reference frame---i.e.\ 
using the observer-centric coordinates---must be smaller than $M_*$. 

In simple spacetimes, we can explicitly see that the local Hawking 
temperatures on surfaces $\lambda = \lambda_{\rm obs}$ determined by 
the condition in Eq.~(\ref{eq:eff-theory}) actually become of order 
$M_*$, so the semi-classical picture is indeed expected to break down 
there.  In these spacetimes, the observer horizons are reduced to the 
stretched horizons defined in Ref.~\cite{Susskind:1993if}.  In de~Sitter 
space, for example, the observer horizon is located at $r = 1/H - 
O(H/M_*^2)$ in the static coordinates when calculated from $p(\tau)$ 
staying at the origin, where $H$ is the Hubble constant.  An important 
point, however, is that unlike the stretched horizon, the definition of 
the observer horizon does not require knowledge of spacetime outside of 
$L_{p_0}$.  This is a desirable feature, as it allows us to construct 
a state without relying on the information in the spacetime region 
outside the one represented by the state.  We also note that the spacetime 
location of the observer horizon, as well as the functional form of 
$\lambda_{\rm obs}(\theta,\phi)$, depends in general on the choice 
of the reference frame $(v^i, \alpha^{[ij]})$.  This is another, 
important difference of the observer horizon from the stretched horizon 
defined in a conventional manner.

We consider that each region of the observer horizon holds ${\cal A}/4 
l_{\rm Pl}^2$ quantum degrees of freedom at the leading order in 
$l_{\rm Pl}^2/{\cal A}$, where ${\cal A}$ is the area of the region.%
\footnote{The number of degrees of freedom is defined as the natural 
 logarithm of the dimension of the corresponding Hilbert space factor. 
 By the leading order, we mean that the number of degrees of freedom 
 is $({\cal A}/4 l_{\rm Pl}^2)\{ 1 + O(l_{\rm Pl}^{2n}/{\cal A}^n) \}$ 
 with $n > 1$.}
This comes from the requirement that the spacetime region ``outside'' 
the observer horizon in the global spacetime picture is reproduced 
by an appropriate reference frame change (complementarity). (See 
Ref.~\cite{Nomura:2013lia} for recent discussions on how this may actually 
work.)  Our picture is such that the degrees of freedom associated with 
the ``outside spacetime'' are entirely in the boundary degrees of freedom 
on the observer horizon.  In fact, as will be discussed in more detail 
in Ref.~\cite{NVW}, the number of the boundary degrees of freedom 
postulated here is sufficient for this purpose because of the holographic 
principle~\cite{'tHooft:1993gx,Susskind:1994vu,Bousso:1999xy}. (In 
the case of a back hole viewed from a distant frame, these degrees 
of freedom are the stretched horizon degrees of freedom.)  An element 
of ${\cal H}$, therefore, may be said to represent a physical state 
of the degrees of freedom in {\it and} on the observer horizon:
\begin{equation}
  0 \,\leq\, \lambda \,\leq\, \lambda_{\rm obs}(\theta,\phi).
\label{eq:boundary-1}
\end{equation}
Note that the bulk and boundary degrees of freedom will in general be 
entangled since the horizon forms by a dynamical process.  Entanglement 
between the two will also be necessary to reconstruct the outside region 
when a relevant reference frame change is made~\cite{VanRaamsdonk:2010pw}.

\subsection{Other ``ends'' of spacetime on {\boldmath $L_{p_0}$}}
\label{subsec:other-ends}

We now discuss other ways in which semi-classical spacetime ceases to 
exist on $L_{p_0}$ along a light ray generating it.  For this purpose, 
we assume that the observer horizon is located sufficiently far away, 
$\lambda_{\rm obs}(\theta,\phi) \rightarrow \infty$.  We argue that 
there are two ways that the light ray may encounter the ``end'' of 
spacetime on $L_{p_0}$ even in this case.

The first possibility is for a light ray to hit a spacetime singularity. 
Consider a null geodesic representing a light ray emitted from $p_0$ 
toward the past in the direction of $(\theta,\phi)$.  Suppose that 
the geodesic encounters a spacetime singularity in the sense that 
it is inextendible beyond some finite value of the affine parameter 
$\lambda_{\rm sing}(\theta,\phi)$ in an inextendible spacetime.  In 
this case, semi-classical spacetime exists only in the region $\lambda 
< \lambda_{\rm sing}(\theta,\phi)$, and we consider that an element 
of ${\cal H}$ represents the physical state of the degrees of freedom 
only in that region.

The other possibility has to do with the behavior of the congruence of 
past-directed light rays emitted from $p_0$.  Assuming the null energy 
condition, $T_{\mu\nu} v^\mu v^\nu \geq 0$ for all null vectors $v^\mu$, 
the expansion of the light rays $\Theta$ satisfies~\cite{Wald:GR}
\begin{equation}
  \frac{\partial \Theta}{\partial \lambda} + \frac{1}{2} \Theta^2 \leq 0.
\label{eq:convergence}
\end{equation}
This implies that the light rays emitted from $p_0$ converge toward the 
past, starting from $\Theta = +\infty$ at $\lambda = 0_+$.

Suppose that a light ray reaches a point where $\Theta = -\infty$ at some 
finite value of the affine parameter $\lambda_{\rm conj}(\theta,\phi)$ 
(before it hits a spacetime singularity).  Such a point is said to be 
conjugate to $p_0$, and signals the failure of the light ray being on the 
boundary of the past of $p_0$~\cite{Wald:GR}.  Specifically, there exists 
a family of timelike causal curves connecting $p_0$ and a point $q$ on 
$L_{p_0}$ with $\lambda > \lambda_{\rm conj}(\theta,\phi)$.  Now, suppose 
semi-classical spacetime exits beyond $\lambda_{\rm conj}(\theta,\phi)$ 
in our framework.  This would contradict the validity of null quantization, 
which we are assuming throughout.  In particular, it would mean that 
a massive particle sent from $q$---which, being on $L_{p_0}$, is at 
an ``equal time'' as $p_0$---can travel backward in time and reach $p_0$ 
from the past (as there exits a timelike causal curve connecting $q$ 
and $p_0$).  We therefore consider that $\Theta = -\infty$ signals the 
end of spacetime, and that an element of ${\cal H}$ only represents 
the region $\lambda < \lambda_{\rm conj}(\theta,\phi)$.

Combining with the possibility of hitting a spacetime singularity 
discussed above, we conclude that an element of ${\cal H}$ represents 
a physical state of the degrees of freedom in the region
\begin{equation}
  0 \,\leq\, \lambda  \,<\, \lambda_{\rm end}(\theta,\phi) 
  \equiv {\rm min} \left\{ \lambda_{\rm sing}(\theta,\phi), 
    \lambda_{\rm conj}(\theta,\phi) \right\},
\label{eq:boundary-2}
\end{equation}
where we have assumed that $\lambda_{\rm obs}(\theta,\phi) > 
\lambda_{\rm end}(\theta,\phi)$.  If a light ray hits the observer 
horizon before it reaches a singularity or a conjugate point, i.e.\ 
$\lambda_{\rm obs}(\theta,\phi) < \lambda_{\rm end}(\theta,\phi)$, 
then spacetime must be terminated there and the boundary degrees of 
freedom must be attached, according to the discussion in the previous 
subsection.

We assume that, unlike the observer horizon, the two-dimensional surface 
determined by $\lambda = \lambda_{\rm end}(\theta,\phi)$ does {\it not} 
hold boundary degrees of freedom.  This corresponds to the hypothesis 
that the evolution of a state can be determined without any information 
from the singularity or the region beyond $\lambda_{\rm conj}(\theta,\phi)$, 
in addition to what is already in the Hamiltonian.  For example, the 
evolution of a big-bang universe is not affected by the ``details'' of 
the big-bang singularity that must be specified beyond the Einstein 
equation.  This conjecture seems to be supported in all the (simple) 
circumstances we have investigated.  Further discussions on this and 
related issues will be given in Ref.~\cite{NVW}.

\subsection{Apparent horizon ``pull-back''}
\label{subsec:apparent}

We have seen that spacetime on the past light cone of $p_0$ 
is extended only until $\lambda$ reaches $\lambda_{\rm obs}$ 
of Section~\ref{subsec:observer-hor} or $\lambda_{\rm end}$ 
of Section~\ref{subsec:other-ends}.  (Here and below, until 
Eq.~(\ref{eq:boundary-final}), we omit the arguments from the boundary 
locations, but it should be remembered that they are functions of 
$\theta$ and $\phi$.)  In the former case, the boundary degrees of 
freedom are attached with the number ${\cal A}/4 l_{\rm Pl}^2$ per 
area ${\cal A}$, while in the latter case, none are attached.  Here 
we discuss a description in which this asymmetry of boundary degrees 
of freedom is dissolved and all the boundaries are treated on equal 
footing for the purpose of counting degrees of freedom.  This description 
is available if the following condition is satisfied:
\begin{equation}
  \lambda_{\rm obs} \leq \lambda_{\rm sing}
 \qquad\mbox{or}\qquad
  \lambda_{\rm conj} \leq \lambda_{\rm sing},
\label{eq:sing-screen}
\end{equation}
i.e.\ a singularity is screened either by the observer horizon or 
conjugate point.  Indeed, in example spacetimes we have investigated, 
this condition is always satisfied, although we do not have a proof 
of it.  Below, we assume that Eq.~(\ref{eq:sing-screen}) is valid, 
and disregard a singularity surface.

Let us define the apparent horizon as a surface on which the expansion 
of the past-directed light rays emitted from $p_0$ first crosses zero:%
\footnote{This definition is different from that in Ref.~\cite{Bousso:2002ju}, 
 where the apparent horizon is defined as a surface on which at least 
 one pair among {\it four} orthogonal null congruences have zero expansion. 
 Here we only consider two directions along $L_{p_0}$.}
\begin{equation}
  \Theta = 0 \quad\mbox{at}\quad \lambda = \lambda_{\rm app}.
\label{eq:lambda_app}
\end{equation}
This implies that $\lambda_{\rm app} < \lambda_{\rm conj}$, since 
$\Theta$ is a monotonically decreasing function of $\lambda$.  Now, 
if $\lambda_{\rm obs} < \lambda_{\rm app}$ for a range of $(\theta,\phi)$, 
then in these directions spacetime ceases to exist at $\lambda = 
\lambda_{\rm obs}$, where a boundary degree of freedom is located 
per area $4 l_{\rm Pl}^2$.  On the other hand, if $\lambda_{\rm app} 
< \lambda_{\rm obs}$ for a range of $(\theta,\phi)$, then there are 
two cases to consider:
\begin{enumerate}
\item $\lambda_{\rm app} < \lambda_{\rm conj} < \lambda_{\rm obs}$ 
--- In this case, spacetime exists only for $\lambda < \lambda_{\rm conj}$. 
The covariant entropy bound then implies that the number of physical 
degrees of freedom in the region $\lambda > \lambda_{\rm app}$ is bounded 
by ${\cal A}/4 l_{\rm Pl}^2$, where ${\cal A}$ is the area of the relevant 
portion of the apparent horizon~\cite{Bousso:1999xy,Bousso:2002ju}.  This 
suggests that these degrees of freedom may be replaced by ${\cal A}/4 
l_{\rm Pl}^2$ boundary degrees of freedom located on the apparent horizon.
\item $\lambda_{\rm app} < \lambda_{\rm obs} < \lambda_{\rm conj}$ 
--- In this case, physical degrees of freedom outside the apparent 
horizon consist of the bulk degrees of freedom in $\lambda_{\rm app} 
< \lambda < \lambda_{\rm obs}$ and the boundary degrees of freedom at 
$\lambda = \lambda_{\rm obs}$.  If the strengthened covariant entropy 
bound of Ref.~\cite{Flanagan:1999jp} applies, then the number of the 
former is bounded by $({\cal A} - {\cal A}_{\rm obs})/4 l_{\rm Pl}^2$, 
while that of the latter is ${\cal A}_{\rm obs}/4 l_{\rm Pl}^2$, where 
${\cal A}$ and ${\cal A}_{\rm obs}$ are the areas of the relevant portions 
of the apparent and observer horizons, respectively.  This suggests that 
physical degrees of freedom in the region $\lambda > \lambda_{\rm app}$ 
may be replaced by ${\cal A}/4 l_{\rm Pl}^2$ boundary degrees of freedom 
on the apparent horizon.  While the strengthened covariant entropy bound 
is known to be violated in some extreme cases, we assume that this 
replacement can always be done in our context.
\end{enumerate}
We thus find that both cases allow for replacing physical degrees of 
freedom in the region $\lambda > \lambda_{\rm app}$ by a quantum degree 
of freedom per area $4 l_{\rm Pl}^2$ on the apparent horizon.  We call 
this replacement procedure {\it apparent horizon pull-back}.

With the apparent horizon pull-back, the structure of the physical 
region represented by an element of ${\cal H}$ can be stated in 
the following simple way.  Spacetime on $L_{p_0}$ exists only for
\begin{equation}
  0 \,\leq\, \lambda \,\leq\, \lambda_B(\theta,\phi) 
  \equiv {\rm min} \left\{ \lambda_{\rm obs}(\theta,\phi), 
    \lambda_{\rm app}(\theta,\phi) \right\}.
\label{eq:boundary-final}
\end{equation}
In addition to the degrees of freedom in the bulk of spacetime, the 
boundary at $\lambda = \lambda_B(\theta,\phi)$ also holds ${\cal A}/4 
l_{\rm Pl}^2$ quantum degrees of freedom (at the leading order in 
$l_{\rm Pl}^2/{\cal A}$), where ${\cal A}$ is the area of the boundary.

\subsection{Horizon decomposition of {\boldmath ${\cal H}$}}
\label{subsec:classify}

So far, we have been discussing the structure of spacetime represented 
by {\it an} element of ${\cal H}$.  The full Hilbert space ${\cal H}$ 
consists of the elements representing ``all possible'' physical 
configurations in ``all possible'' spacetimes, as viewed from the 
reference frame.  What do we really mean by that?  In other words, 
what is the structure of ${\cal H}$ concretely?

To address this question, let us adopt the apparent-horizon pulled-back 
description, discussed in the previous subsection.  We now group the 
elements that have the same boundary $\partial{\cal M}$, and denote the 
Hilbert space spanned by these elements by ${\cal H}_{\partial{\cal M}}$.%
\footnote{The ${\cal H}_{\partial{\cal M}}$ here is the same 
 as what is denoted by ${\cal H}_{\cal M}$ in earlier papers 
 Refs.~\cite{Nomura:2011rb,Nomura:2012zb,Nomura:2012cx}.}
The general definition of the boundary being the same is not obvious 
to give explicitly.  One possible definition, which seems to work 
if the boundary is within the coordinate horizon $g_{\tau\tau} = 0$, 
is given as follows.  Consider the induced metric on the boundary 
$\lambda = \lambda_B(\theta,\phi)$ with the arguments being the 
observer-centric angular variables:
\begin{equation}
  h_{XY}(\theta,\phi) = \frac{\partial \lambda_B}{\partial X} 
    \frac{\partial \lambda_B}{\partial Y} g_{\lambda\lambda} 
    + \frac{\partial \lambda_B}{\partial X} g_{\lambda Y} 
    + \frac{\partial \lambda_B}{\partial Y} g_{\lambda X} 
    + g_{XY},
\label{eq:induced}
\end{equation}
where $X,Y = \theta,\phi$, and $g_{\lambda\lambda}$, $g_{\lambda X}$, and 
$g_{XY}$ are spacetime metric components in the observer-centric coordinate 
system, evaluated at $\tau = 0$ and $\lambda = \lambda_B(\theta,\phi)$. 
We regard two boundaries as the same if the induced metrics on them are 
{\it explicitly} identical, i.e.\ all the $h_{XY}$'s ($X,Y = \theta,\phi$) 
take the identical functional forms with respect to $\theta$, $\phi$.%
\footnote{It is not entirely clear if there is no additional condition 
 for the boundaries being the same; for example, we might have to 
 require $\lambda_B(\theta,\phi)$ to be the same in addition to 
 $h_{XY}(\theta,\phi)$.  Here we postulate that the identity of 
 $h_{XY}(\theta,\phi)$ is sufficient, and proceed with it.}

This definition reflects the fact that our description of physics is 
``special relativistic'' or ``as viewed from the reference frame.'' 
For example, a spacetime 2-surface is regarded as different boundaries 
when described from two different reference frames which are rotated with 
respect to with each other (unless the surface is spherically symmetric 
around $p_0$).  This implies that depending on the choice of the 
reference frame, the identical physical configuration in spacetime 
can belong to different Hilbert subspaces ${\cal H}_{\partial{\cal M}}$. 
An operator corresponding to rotating the reference frame then transforms 
an element of a subspace into that of another.  Note that here we are 
talking about a state $\ket{\psi_i}$ in ${\cal H} \subset {\cal H}_{\rm QG}$, 
which may be viewed as representing a physical state relative to clock 
degrees of freedom.  The ``full'' quantum state (i.e.\ the multiverse 
state) $\ket{\Psi} \subset {\cal H}_{\rm phys}$ obtained after imposing 
the constraints in Eq.~(\ref{eq:Poincare}) is, of course, invariant 
under such a rotation (guaranteeing that there is no absolute frame 
in the universe).

Now, the elements of ${\cal H}_{\partial{\cal M}}$ represent all 
possible physical configurations in all possible spacetimes (or null 
slices of spacetimes) that share the same boundary $\partial{\cal M}$ 
as defined above.  Let us denote the Hilbert space factors of 
${\cal H}_{\partial{\cal M}}$ corresponding to the bulk and boundary 
degrees of freedom by ${\cal H}_{\partial{\cal M},\,{\rm bulk}}$ and 
${\cal H}_{\partial{\cal M},\,B}$, respectively:
\begin{equation}
  {\cal H}_{\partial{\cal M}} 
    = {\cal H}_{\partial{\cal M},\,{\rm bulk}} 
    \otimes {\cal H}_{\partial{\cal M},\,B},
\label{eq:H_M-decomp}
\end{equation}
where the direct product structure arises from the locality 
hypothesis in our framework.  According to the covariant entropy 
bound~\cite{Bousso:1999xy}, the dimension of the Hilbert space 
factor ${\cal H}_{\partial{\cal M},\,{\rm bulk}}$ is bounded 
by the area of the boundary ${\cal A}_{\partial {\cal M}}$ as 
${\rm dim}\,{\cal H}_{\partial{\cal M},\,{\rm bulk}} \leq 
\exp({\cal A}_{\partial {\cal M}}/4l_{\rm Pl}^2)$.  On the 
other hand, by construction the dimension of the boundary 
factor is ${\rm dim}\,{\cal H}_{\partial{\cal M},\,B} = 
\exp({\cal A}_{\partial {\cal M}}/4l_{\rm Pl}^2)$.  Therefore, 
we find
\begin{equation}
  {\rm dim}\,{\cal H}_{\partial{\cal M}} 
  = {\rm dim}\,{\cal H}_{\partial{\cal M},\,{\rm bulk}}  \times 
    {\rm dim}\,{\cal H}_{\partial{\cal M},\,B} 
  \leq \exp\left(\frac{{\cal A}_{\partial {\cal M}}}{2 l_{\rm Pl}^2}\right).
\label{eq:H_M-dimension}
\end{equation}
Note that this includes arbitrary fluctuations of spacetimes as well 
as arbitrary configurations of matter (which are related by Einstein's 
equation with each other) that keep the boundary fixed, namely with 
$h_{XY}(\theta,\phi)$ held fixed.%
\footnote{Recently, the analysis above has been significantly refined 
 in Ref.~\cite{Nomura:2013lia}, which claims that for physical states 
 the relevant space is given by ${\cal H}_{\partial{\cal M}}$ with 
 ${\rm dim}\,{\cal H}_{\partial{\cal M}} = \exp({\cal A}_{\partial 
 {\cal M}} /4 l_{\rm Pl}^2)$ (at least at leading order in an $l_{\rm Pl}^2 
 / {\cal A}_{\partial {\cal M}}$ expansion in the exponent), which is 
 much smaller than $\exp({\cal A}_{\partial {\cal M}} / 2 l_{\rm Pl}^2)$ 
 appearing in the last expression in Eq.~(\ref{eq:H_M-dimension}).  This 
 is possible because the contribution from the bulk region is in general 
 tiny $\approx O({\cal A}^n/l_{\rm Pl}^{2n})$ ($n < 1$)~\cite{'tHooft:1993gx} 
 for physically realizable states, and hence can be neglected at the 
 leading order.  In fact, when $\partial {\cal M}$ is the observer 
 horizon, we find that $\ln {\rm dim}\,{\cal H}_{\partial{\cal M}}$ 
 for physical states is saturated (at the leading order in $l_{\rm Pl}^2 
 / {\cal A}_{\partial {\cal M}}$) by the {\it entropy of a vacuum}---the 
 logarithm of the number of possible independent ways in which quantum 
 field theory on a fixed classical spacetime background can emerge in 
 a full quantum theory of gravity~\cite{Nomura:2013lia}.}

The complete spacetime part of the Hilbert space ${\cal H}$ is then given 
by the direct sum of the Hilbert subspaces ${\cal H}_{\partial{\cal M}}$ 
for different $\partial{\cal M}$'s:
\begin{equation}
  {\cal H} = \bigoplus_{\partial{\cal M}} {\cal H}_{\partial{\cal M}},
\label{eq:H-final}
\end{equation}
where the direct sum runs over $\partial{\cal M} = \{ h_{XY}(\theta,\phi) 
\}$.  We call the expression of this form the {\it horizon decomposition} 
of ${\cal H}$.  In general, what $\partial{\cal M}$'s are included 
in the decomposition of the complete Hilbert space ${\cal H}$ cannot 
be determined by the low energy consideration alone.  For instance, 
some spacetimes such as stable (not cosmological) de~Sitter space 
may be unrealistic mathematical idealizations and may not appear in 
the underlying full quantum theory of gravity.  In practice, however, 
we may include only $\partial{\cal M}$'s that are relevant to the 
problem under consideration (the ones relevant for the clock degrees 
of freedom), and that is sufficient.  For discussions of this issue 
in cosmology, especially in the eternally inflating multiverse, see 
Ref.~\cite{Nomura:2012zb}.

\subsection{Spacelike quantization}
\label{subsec:spacelike}

Finally, we discuss briefly if there is a way to use spacelike 
hypersurfaces, rather than null hypersurfaces, to quantize the system. 
Such a spacelike quantization would avoid technical subtleties associated 
with null quantization, for example, non-commutativity of field operators 
at different points in a same angular direction (see footnote~\ref{ft:null}).

One possibility is simply to ``round'' the light cone $L_{p_0}$ slightly 
to make an equal-time hypersurface spacelike.  We can do this while 
keeping the boundary $\partial{\cal M}$ fixed.  An advantage of this 
procedure is that the structure of the Hilbert space is unchanged from 
that in Eqs.~(\ref{eq:H_M-decomp}~--~\ref{eq:H-final}).  This is because 
the future-directed ingoing light sheet of $\partial{\cal M}$ (a portion 
of $L_{p_0}$ bounded by $\partial{\cal M}$) is complete (ending at 
the caustic at $p_0$), so that the spacelike projection theorem of 
Ref.~\cite{Bousso:1999xy} applies.  In a sense, our null quantization 
may be viewed as a limit of the spacelike quantization discussed here 
(although the limit is not completely smooth).

Another possibility is to adopt an ``intrinsically spacelike'' construction. 
Specifically, we may follow a similar construction to our covariant 
Hilbert space using spacelike geodesics attached to the local Lorentz 
frame at $p_0$ (e.g.\ with the affine parameters taken to agree with 
the radial coordinate in the infinitesimal vicinity of $p_0$), instead 
of null geodesics (light rays).  In particular, we may define acceleration 
parameter $A$ and the observer horizon similarly in a static situation. 
This construction corresponds to taking different linear combinations 
of the constraint operators ${\cal P}^\mu({\bf x})$ as $H$, $P_i$, 
$J_{[ij]}$, and $K_i$ (see discussion in Section~\ref{sec:QHS}).  The 
validity of this approach or its relation to the null quantization 
presented in this letter is not fully clear.  We plan to study this 
possibility further in Ref.~\cite{NVW}.

\section{Implications on (No) Firewalls}
\label{sec:firewall}

The complementarity picture adopted in our framework is such that 
the spacetime region outside the observer horizon of $p_0$---which 
appears (only) after a reference frame change---is reproduced from the 
boundary degrees of freedom located on the horizon.  Such a picture has 
recently been challenged by AMPS~\cite{Almheiri:2012rt,Almheiri:2013hfa} 
(see also~\cite{Braunstein:2009my}), who claim that the smoothness of 
the horizon (the equivalence principle) together with the semi-classical 
nature of spacetime at length scales larger than $l_*$ is fundamentally 
incompatible with unitarity of quantum mechanics.  This issue is 
still under debate, and we do not directly address it here.  Rather, 
we ask what implications the current framework will have if the 
issue is somehow resolved, for example along the lines of 
Refs.~\cite{Nomura:2012ex,Nomura:2013lia} (see also~\cite{Verlinde:2012cy} 
for a similar construction).  In particular, we discuss how the 
degrees of freedom which in a distant picture can be viewed as 
associated with the black hole may appear in an infalling picture.

Let us recall the ``firewall'' argument by AMPS.  Consider three subsystems 
of an old black hole system.  In a distant frame, these are taken to be 
(1) $R$: early/distant Hawking modes, (2) $B$: outgoing modes localized 
near outside of a (small) patch of the horizon, and (3) $A$: a subsystem 
of the degrees of freedom composing the stretched horizon that is entangled 
with the modes in $B$.%
\footnote{Following the convention in recent literature, we 
 have changed the symbols denoting various modes.  Specifically, 
 $R$, $B$, and $A$ here correspond to $A$, $B$, and $C$ in 
 Refs.~\cite{Almheiri:2012rt,Nomura:2012ex}.}
In an infalling frame, the interpretation of $A$ (but not of $R$ or 
$B$) changes, although it still represents the same degrees of freedom 
(complementarity): (3)$^\prime$ $A$: modes inside the horizon that are 
partner modes of $B$.  Now, unitarity implies that for an old black hole 
the entropy of the distant modes decreases as the near modes get out
$S_{BR} < S_R$, where $S_X$ represents the von~Neumann entropy of system 
$X$.  On the other hand, the equivalence principle applied to a freely 
falling observer says $S_{AB} = 0$, implying $S_{ABR} = S_R$.  These 
two relations contradict strong subadditivity of entropy $S_{AB} + S_{BR} 
\geq S_B + S_{ABR}$, since they lead to $S_B < 0$ if both are true. 
From this, AMPS conclude that one must abandon either unitarity of 
a black hole formation/evaporation process (with physics outside 
the stretched horizon well described by a semi-classical theory) 
or the equivalence principle.

In Ref.~\cite{Nomura:2012ex}, it was argued that this conclusion may 
be avoided since an infalling vacuum may be realized in multiple different 
ways at the microscopic level because of large microscopic degrees of 
freedom of the black hole.  Specifically, let us write the quantum state 
of an old black hole and early radiation as viewed from a distant frame as
\begin{equation}
  \ket{\Phi} = \sum_i c_i \ket{R_i} \ket{{\rm BH}_i},
\label{eq:BH-state}
\end{equation}
where $\ket{R_i}$ represent states for $R$ in the basis in which 
$\ket{R_i}$'s have well-defined phase space configurations, while 
$\ket{{\rm BH}_i}$ are those for the rest of the system, which includes 
$B$, $A$, and the other modes of the black hole than $A$.  By expanding 
$\ket{{\rm BH}_i}$ in terms of different microscopic vacuum states 
$\ket{V_a}$ as $\ket{{\rm BH}_i} = \sum_a d_{ia} \ket{V_a}$, the combined 
black hole and radiation state $\ket{\Phi}$ can be written as
\begin{equation}
  \ket{\Phi} = \sum_{i,a} c_i\, d_{ia} \ket{R_i} \ket{V_a} 
    \equiv \sum_n \omega_n |\tilde{\Phi}_n \rangle,
\label{eq:BH-state-2}
\end{equation}
where $n \equiv \{ i, a \}$ and $|\tilde{\Phi}_n \rangle \equiv \ket{R_i} 
\ket{V_a}$.  Unitarity of quantum mechanics says that $\ket{\Phi}$ 
satisfies $S_{BR} < S_R$, implying that $S_{AB} = 0$ cannot be true. 
This, however, does not necessarily mean that general relativity is 
incorrect.  The validity of the equivalence principle requires that 
the $AB$ system is maximally entangled in each classical branch:
\begin{equation}
  \tilde{S}^{(n)}_{AB} = 0,
\label{eq:ent-rel-3}
\end{equation}
where $\tilde{S}^{(n)}_{AB}$ is the {\it branch world entropy} of subsystem 
$AB$ in classical world $n$~\cite{Nomura:2012ex}, i.e.\ the von~Neumann 
entropy of $AB$ calculated using the state $|\tilde{\Phi}_n \rangle$, rather 
than $\ket{\Phi}$.  The point is that relations in Eq.~(\ref{eq:ent-rel-3}) 
are {\it not} incompatible with the unitarity relation $S_{BR} < S_R$.

The argument described above addresses some aspects of the firewall paradox, 
if not all.  A crucial element there is the existence of (exponentially) 
many possible ways in which the infalling vacuum is encoded in the 
stretched horizon modes at the microscopic level, which we identify 
as the origin of the Bekenstein-Hawking entropy.  We now ask the question 
(regardless of the full validity of the scenario described above):\ 
if there are indeed exponentially many possible vacuum states 
$|\tilde{\Phi}_n \rangle$ in a distant picture, what is the description 
of them in an infalling frame?  In Ref.~\cite{Nomura:2011rb}, complementarity 
was postulated to be unitary transformations acting on ${\cal H}_{\rm QG}$ 
(more fundamentally, a change of the clock degrees of freedom accompanied 
by the action of an appropriate combination of the Poincar\'{e} operators 
$H, P_i, J_{[ij]}, K_i$ on relative states $\ket{\psi_i}$).  This may 
not be true; for example, recent analyses in Ref.~\cite{Verlinde:2012cy} 
find that a complementarity transformation is state-dependent, rather 
than unitary.  Regardless, if a complementarity transformation does not 
eliminate (microscopic) degrees of freedom, different $|\tilde{\Phi}_n 
\rangle$'s in a distant frame must be mapped into different states in 
an infalling frame.  However, in the infalling frame, all these states 
must represent an infalling vacuum in spacetime where there is no black 
hole horizon.  Where do the necessary degrees of freedom to discriminate 
these states come from?

We conjecture that they come from boundary degrees of freedom on the 
observer horizon (more precisely a part of the observer horizon associated 
with the existence of the black hole) of the infalling reference frame. 
Remember that the location of the observer horizon depends on the 
choice of the reference frame, in particular the spacetime location 
$q^\mu$ and velocity $v^i$ of the reference point $p_0$.  If $p_0$ 
is moving slowly at a location far from the black hole
\begin{equation}
  v_p \ll 1,
\qquad
  r_p \gg R_S,
\label{eq:BH-p0}
\end{equation}
then the observer horizon associated with the black hole agrees with the 
conventional stretched horizon, located at
\begin{equation}
  r_{\rm obs} = R_S + O\left( \frac{1}{R_S M_*^2} \right),
\end{equation}
where $v_p = |v_i|$, $r_p$ is the radial component of $q^\mu$ in the 
Schwarzschild coordinates, and $R_S$ is the Schwarzschild radius~\cite{NVW}.%
\footnote{The part of the observer horizon associated with the black 
 hole can be determined by the values of $\lambda_{\rm obs}$.  For 
 example, in Schwarzschild-de~Sitter spacetime with $p_0$ located at 
 $r_p \ll H^{-1}$ outside the black hole, the values of $\lambda_{\rm obs}$ 
 are of $O(r_p)$ in a small angular region $\varDelta\Omega$ while of 
 $O(H^{-1})$ in the rest.  The observer horizon in $\varDelta\Omega$ 
 is then associated with the black hole, while the rest with the de~Sitter 
 horizon.  For more detailed discussions, see Ref.~\cite{NVW}.}
When $p_0$ enters into the black hole (i.e.\ $r_p$ decreases passed 
$R_S$), the observer horizon recedes so that $p_0$ does not hit the 
observer horizon (until it reaches the singularity when $p_0$ collides 
with the observer horizon).  We expect that until $r_p$ becomes much 
smaller than $R_S$, the area of the observer horizon associated with 
the black hole is of order $R_S^2$, so that there are enough degrees 
of freedom on the horizon (of order the black hole entropy) to which 
different $|\tilde{\Phi}_n \rangle$'s are mapped into.

In summary, we conjecture that a complementarity transformation provides 
the following mapping at the microscopic level:
\begin{itemize}
\item[] {\bf Distant view:} Different microscopic encodings of the 
 infalling vacuum (and a fallen object) in the degrees of freedom of 
 the black hole stretched horizon (which comprises a part of the observer 
 horizon of the distant reference frame)
 \\
 $\Longleftrightarrow$\\
 {\bf Infalling view:} The infalling vacuum (and a falling object) in 
 spacetime with different microstates for the observer horizon of the 
 infalling reference frame
\end{itemize}
In the infalling view, unitarity of the entire quantum state $\ket{\Phi}$ 
is ensured by entanglement between early radiation modes and different 
``observer horizon worlds,'' each of which behaves as a different, 
decohered classical world.  The equivalence principle is intact 
if the structure of the states around $p_0$ is correctly Minkowski 
vacuum-like in these classical worlds.  To show that it is actually 
the case, however, requires a machinary beyond the one presented here.

\section*{Acknowledgments}

This work was supported in part by the Director, Office of Science, Office 
of High Energy and Nuclear Physics, of the US Department of Energy under 
Contracts DE-FG02-05ER41360 and DE-AC02-05CH11231, by the National Science 
Foundation under grants PHY-0855653 and DGE-1106400, and by the Simons 
Foundation grant 230224.


\begin{thebibliography}{99}

\bibitem{Preskill:1992tc}
For reviews, see e.g.\
J.~Preskill,
%``Do black holes destroy information?,''
in {\it Blackholes, Membranes, Wormholes and Superstrings},
ed.\ S.~Kalara and D.~V.~Nanopoulos (World Scientific, Singapore, 1993) p.~22
[hep-th/9209058];
%%CITATION = HEP-TH/9209058;%%
%\bibitem{Susskind:2005js}
L.~Susskind and J.~Lindesay,
{\it An Introduction to Black Holes, Information and the String Theory 
 Revolution: The Holographic Universe}
(World Scientific, Singapore, 2005).

\bibitem{Susskind:1993if}
L.~Susskind, L.~Thorlacius and J.~Uglum,
%``The Stretched Horizon And Black Hole Complementarity,''
Phys.\ Rev.\  D {\bf 48}, 3743 (1993)
[arXiv:hep-th/9306069].
%%CITATION = PHRVA,D48,3743;%%

\bibitem{Stephens:1993an}
C.~R.~Stephens, G.~'t Hooft and B.~F.~Whiting,
%``Black hole evaporation without information loss,''
Class.\ Quant.\ Grav.\  {\bf 11}, 621 (1994)
[arXiv:gr-qc/9310006].
%%CITATION = CQGRD,11,621;%%

\bibitem{Nomura:2011rb}
Y.~Nomura,
%``Quantum Mechanics, Spacetime Locality, and Gravity,''
Found.\ Phys.\  {\bf 43}, 978 (2013)
[arXiv:1110.4630 [hep-th]].
%%CITATION = ARXIV:1110.4630;%%

\bibitem{Nomura:2011dt}
Y.~Nomura,
%``Physical Theories, Eternal Inflation, and Quantum Universe,''
JHEP {\bf 11}, 063 (2011)
[arXiv:1104.2324 [hep-th]].
%%CITATION = ARXIV:1104.2324;%%

\bibitem{Almheiri:2012rt}
A.~Almheiri, D.~Marolf, J.~Polchinski and J.~Sully,
%``Black Holes: Complementarity or Firewalls?,''
JHEP {\bf 02}, 062 (2013)
[arXiv:1207.3123 [hep-th]].
%%CITATION = ARXIV:1207.3123;%%

\bibitem{NVW}
Y.~Nomura, J.~Varela and S.~J.~Weinberg,
in preparation.

\bibitem{Dirac-book}
P.~A.~M.~Dirac,
{\it Lectures on Quantum Mechanics},
(Belfer Graduate School of Science, Yeshiva University, New York, 1964).

\bibitem{DeWitt:1967yk}
B.~S.~DeWitt,
%``Quantum Theory of Gravity. 1. The Canonical Theory,''
Phys.\ Rev.\  {\bf 160}, 1113 (1967).
%%CITATION = PHRVA,160,1113;%%

\bibitem{Nomura:2012zb}
Y.~Nomura,
%``The Static Quantum Multiverse,''
Phys.\ Rev.\ D {\bf 86}, 083505 (2012)
[arXiv:1205.5550 [hep-th]].
%%CITATION = ARXIV:1205.5550;%%

\bibitem{Nomura:2012cx}
Y.~Nomura, J.~Varela and S.~J.~Weinberg,
%``Black Holes, Information, and Hilbert Space for Quantum Gravity,''
Phys.\ Rev.\ D {\bf 87}, 084050 (2013)
[arXiv:1210.6348 [hep-th]].
%%CITATION = ARXIV:1210.6348;%%

\bibitem{Nomura:2013lia}
Y.~Nomura and S.~J.~Weinberg,
%``The Entropy of a Vacuum: What Does the Covariant Entropy Count?,''
arXiv:1310.7564 [hep-th].
%%CITATION = ARXIV:1310.7564;%%

\bibitem{'tHooft:1993gx}
G.~'t Hooft,
%``Dimensional reduction in quantum gravity,''
arXiv:gr-qc/9310026.
%%CITATION = GR-QC/9310026;%%

\bibitem{Susskind:1994vu}
L.~Susskind,
%``The World As A Hologram,''
J.\ Math.\ Phys.\  {\bf 36}, 6377 (1995)
[arXiv:hep-th/9409089].
%%CITATION = JMAPA,36,6377;%%

\bibitem{Bousso:1999xy}
R.~Bousso,
%``A Covariant Entropy Conjecture,''
JHEP {\bf 07}, 004 (1999)
[arXiv:hep-th/9905177].
%%CITATION = JHEPA,9907,004;%%

\bibitem{VanRaamsdonk:2010pw}
See, e.g.,
M.~Van Raamsdonk,
%``Building up spacetime with quantum entanglement,''
Gen.\ Rel.\ Grav.\  {\bf 42}, 2323 (2010)
[Int.\ J.\ Mod.\ Phys.\ D {\bf 19}, 2429 (2010)]
[arXiv:1005.3035 [hep-th]].
%%CITATION = ARXIV:1005.3035;%%

\bibitem{Wald:GR}
See, e.g.,
R.~M.~Wald,
{\it General Relativity}
(The University of Chicago Press, Chicago, 1984).

\bibitem{Bousso:2002ju}
R.~Bousso,
%``The Holographic principle,''
Rev.\ Mod.\ Phys.\  {\bf 74}, 825 (2002)
[hep-th/0203101].
%%CITATION = HEP-TH/0203101;%%

\bibitem{Flanagan:1999jp}
\'{E}.~\'{E}.~Flanagan, D.~Marolf and R.~M.~Wald,
%``Proof of classical versions of the Bousso entropy bound and of the generalized second law,''
Phys.\ Rev.\ D {\bf 62}, 084035 (2000)
[hep-th/9908070].
%%CITATION = HEP-TH/9908070;%%

\bibitem{Almheiri:2013hfa}
A.~Almheiri, D.~Marolf, J.~Polchinski, D.~Stanford and J.~Sully,
%``An Apologia for Firewalls,''
JHEP {\bf 09}, 018 (2013)
[arXiv:1304.6483 [hep-th]];
%%CITATION = ARXIV:1304.6483;%%
%\bibitem{Marolf:2013dba}
D.~Marolf and J.~Polchinski,
%``Gauge/Gravity Duality and the Black Hole Interior,''
Phys.\ Rev.\ Lett.\  {\bf 111}, 171301 (2013)
[arXiv:1307.4706 [hep-th]].
%%CITATION = ARXIV:1307.4706;%%

\bibitem{Braunstein:2009my}
S.~L.~Braunstein,
%``Entangled black holes as ciphers of hidden information,''
arXiv:0907.1190v1 [quant-ph];
%%CITATION = ARXIV:0907.1190;%%
%\bibitem{Mathur:2009hf}
S.~D.~Mathur,
%``The Information paradox: A Pedagogical introduction,''
Class.\ Quant.\ Grav.\  {\bf 26}, 224001 (2009)
[arXiv:0909.1038 [hep-th]].
%%CITATION = ARXIV:0909.1038;%%

\bibitem{Nomura:2012ex}
Y.~Nomura and J.~Varela,
%``A Note on (No) Firewalls: The Entropy Argument,''
JHEP {\bf 07}, 124 (2013)
[arXiv:1211.7033 [hep-th]].
%%CITATION = ARXIV:1211.7033;%%

\bibitem{Verlinde:2012cy}
E.~Verlinde and H.~Verlinde,
%``Black Hole Entanglement and Quantum Error Correction,''
JHEP {\bf 10}, 107 (2013)
[arXiv:1211.6913 [hep-th]].
%%CITATION = ARXIV:1211.6913;%%
%\bibitem{Papadodimas:2012aq}
K.~Papadodimas and S.~Raju,
%``An Infalling Observer in AdS/CFT,''
JHEP {\bf 10}, 212 (2013)
[arXiv:1211.6767 [hep-th]];
%%CITATION = ARXIV:1211.6767;%%
%\bibitem{Papadodimas:2013jku}
%K.~Papadodimas and S.~Raju,
%``State-Dependent Bulk-Boundary Maps and Black Hole Complementarity,''
arXiv:1310.6335 [hep-th].
%%CITATION = ARXIV:1310.6335;%%

\end{thebibliography}
\end{document}